\documentclass[aps,pra,twocolumn,showpacs,superscriptaddress]{revtex4-1}
\usepackage[colorlinks,linkcolor=blue,anchorcolor=blue,citecolor=blue]{hyperref}
\usepackage{graphicx,amsmath,amssymb}
\usepackage[usenames]{color}
\bibpunct{[}{]}{,}{s}{}{;}

\begin{document}

\title{Hidden Single-Qubit Topological Phase Transition without Gap Closing in Anisotropic Light-Matter Interactions
}

\author{Zu-Jian Ying }
\email{yingzj@lzu.edu.cn}
%\email{zujianying@yahoo.com}
\affiliation{School of Physical Science and Technology, Lanzhou University, Lanzhou 730000, China}
%\affiliation{CNR-SPIN and Dipartimento di Fisica ``E. R. Caianiello'', Universit\`a di Salerno, 84084 Fisciano, Italy}

\begin{abstract}
Conventionally the occurrence of topological phase transitions (TPTs) requires gap closing, whereas there are also unconventional cases without need of gap closing. Although traditionally TPTs lie in many-body systems in condensed matter, both cases of TPTs may find analogs in few-body systems. Indeed, the ground-state node number provides a topological classification for single-qubit systems. While the no-node theorem of spinless systems is shown to also restrict the fundamental quantum Rabi model in light-matter interactions, it is demonstrated that the limitation of the no-node theorem can be broken not only in a small counter-rotating term (CRT) but also in the large-CRT regime, which striates a rich phase diagram with different TPTs. While these transitions are mostly accompanied with gap closing and parity reversal, a hidden node-phase transition is revealed that has neither gap closing nor parity change, which turns out to be an analog of the unconventional TPT in condensed matter. A hysteresis sign for the unconventional TPT is unveiled via the transition from amplitude squeezing to phase squeezing in the gapped phase. The imprints in the Wigner function are also addressed. The clarified mechanisms provide some special insights for the subtle role of the CRT.
\end{abstract}
\pacs{ }
\maketitle

%\date{\today}

\section{Introduction}

Light-matter interactions play a ubiquitous role in our physical world and
the past decade has witnessed both extraordinary experimental progresses
\cite{Diaz2019RevModPhy,Kockum2019NRP} and tremendous theoretical efforts
\cite{Braak2011,Boite2020,Liu2021AQT} on the investigations of light-matter
interactions in the frontiers of modern quantum physics and quantum
technologies. Indeed, the continuing experimental enhancements \cite%
{Diaz2019RevModPhy,Wallraff2004,Gunter2009,Niemczyk2010,Peropadre2010,FornDiaz2017, Ciuti2005EarlyUSC,Aji2009EarlyUSC,Forn-Diaz2010,Scalari2012,Xiang2013,Yoshihara2017NatPhys,Kockum2017}
of the interaction strength have brought a new era with ultra-strong\cite%
{Diaz2019RevModPhy,Wallraff2004,Gunter2009,Niemczyk2010,Peropadre2010,FornDiaz2017, Forn-Diaz2010,Scalari2012,Xiang2013,Yoshihara2017NatPhys,Kockum2017}
and even deep-strong couplings.\cite{Yoshihara2017NatPhys,Bayer2017DeepStrong}
The milestone work \cite{Braak2011} of revealing the Braak integrability for the quantum Rabi model (QRM),
which is a most fundamental model of light-interactions, has stimulated
intense theoretical studies. \cite%
{Boite2020,Wolf2012,FelicettiPRL2020,Felicetti2018-mixed-TPP-SPP,Felicetti2015-TwoPhotonProcess,Simone2018, Solano2011,Irish2014,Irish2017,PRX-Xie-Anistropy,Batchelor2015,XieQ-2017JPA,Hwang2015PRL,Bera2014Polaron, Ying2015,LiuM2017PRL,Ying-2018-arxiv,Ying-2021-AQT,CongLei2017,CongLei2019,Ying2020-nonlinear-bias,Liu2021AQT,ChenQH2012,e-collpase-Duan-2016,ZhangYY2016,ZhengHang2017,PengJie2019,Liu2015,Ashhab2013, ChenGang2012,FengMang2013,Eckle-2017JPA,Casanova2018npj,HiddenSymMangazeev2021,HiddenSymLi2021,HiddenSymBustos2021,Garbe2020}

As well-known, under the rotating-wave approximation the QRM \cite{rabi1936}
reduces to the Jaynes-Cummings model (JCM),\cite{JC-model} which applies
for weak/strong couplings or around resonance but fails in ultra-strong and
deep-strong regimes. \cite%
{Irish2014,Ying2015,Liu2021AQT,Kockum2019NRP,PRX-Xie-Anistropy}
Nevertheless, the invalidity of the approximation does not exclude possible
ultra-strong or deep-strong coupling of the JCM itself.\cite%
{Ulstrong-JC-1,Ulstrong-JC-2} These two models are also fundamental building
blocks for quantum information and quantum computation \cite%
{Diaz2019RevModPhy,Romero2012,Stassi2020QuComput,Stassi2018,Macri2018} and
closely connected to models in condense matter \cite{Kockum2019NRP} and
relativistic systems. \cite{Bermudez2007} A full understanding of their
difference, namely the counter-rotating term (CRT), is a main concern both
experimentally\cite{Forn-Diaz2010,Pietikainen2017,Yimin2018} and
theoretically.\cite{PRX-Xie-Anistropy,LiuM2017PRL,Ying-2021-AQT} The
convenient model to study the role of the CRT is the anisotropic QRM which
connects the QRM and the JCM by continuously tuning the CRT via the anisotropy of
the coupling.\cite{PRX-Xie-Anistropy,LiuM2017PRL,Ying-2021-AQT}

One of the most fascinating phenomena in enhancing the coupling is possible
onset of few-body phase transitions.\cite{Liu2021AQT,Ashhab2013,Ying2015,Hwang2015PRL,Ying2020-nonlinear-bias,Ying-2021-AQT,LiuM2017PRL,Hwang2016PRL}
When phase transitions
traditionally lie in thermodynamic limit in condensed matter, the QRM
possesses a few-body quantum phase transition in the low frequency limit.\cite{Ashhab2013,Ying2015,Hwang2015PRL}
On the other hand, it was also suggested that
whether the transition should be termed quantum or not is a matter of taste
by considering the negligible quantum fluctuations in the photon vacuum
state.\cite{Irish2017} Moreover, a single-qubit system can even exhibit
multicriticalities either with different patterns of symmetry breaking\cite{Ying2020-nonlinear-bias}
or with the parity symmetry preserved.\cite{Ying-2021-AQT} Interestingly, the anisotropic QRM exhibits a universality
of criticality in the scaling of critical exponents.\cite{LiuM2017PRL}
However, such a universality holds only under the condition of low frequency
limit, while at finite frequencies the universality breaks down and
diversity arises.\cite{Ying-2021-AQT}

Surprisingly, among the dominant diversity a new universality classification
can be extracted from the topological feature of the ground state with
different node numbers and the series transitions emerging at finite
frequencies were found \cite{Ying-2021-AQT} to be analogs of symmetry-protected topological
phase transitions (TPTs) in condensed matter\cite{Topo-KT-transition,Topo-KT-NoSymBreak,Topo-Haldane-1,Topo-Haldane-2,Topo-Wen}
which are essentially different
from the Laudau class of phase transitions with symmetry breaking.\cite{Landau-theory}
These emerging topological transitions in bridging the QRM and the JCM occur
with gap closing and re-opening just as the conventional TPTs in condensed
matter. Although gap closing is a necessary condition for the conventional
TPTs of non-degenerate states, there are also some
unconventional TPTs in condensed matter that occur without gap closing in
some particular situations, such as in the presence of a strong
electron-electron interaction in the quantum spin Hall effect\cite%
{Amaricci-2015-no-gap-closing} or in the presence of disorder with Berry
curvature separation in the quantum anomalous Hall effect.\cite{Xie-QAH-2021}
Since the single-qubit system of the anisotropic QRM can exhibit TPTs with
gap closing as in the many-body systems,\cite{Ying-2021-AQT} one may wonder
whether a single-qubit system can also have any analog of the unconventional
TPTs without gap closing.

To seek such a possibility we study in the present work the ground state and
the first excitation gap of the anisotropic QRM in the full parameter space
of coupling and anisotropy. We find that the limitation of the no-node
theorem is broken not only in the small-CRT regime between the QRM and the
JCM but also in the large-CRT regime beyond the QRM. While the TPTs in the
both regimes are conventional ones with gap closing, we find there is a
hidden transition between the two regimes that turns out to be an
unconventional TPT without gap closing. We clarify the underlying mechanisms
and show that the node in the conventional TPTs appears around the origin
thus changing the parity, while the node in the unconventional TPT comes from
the infinity thus keeping the parity unaffected. In the gapped phase we also reveal a transition
from amplitude squeezing to phase squeezing which can be regarded as a hysteresis sign for the unconventional TPT.

The paper is organized as follows. Section \ref{Sect-Model} introduces the
anisotropic QRM and the transform to quadrature representation. In Section %
\ref{Sect-NoNode-Rabi} the no-node theorem is established for the QRM.
Section \ref{Sect-Top-with-gap-closing} shows the conventional TPTs and
Section \ref{Sect-Top-without-gap-closing} reveals the unconventional TPT.
The AS/PS transition is addressed in Section \ref{Sect-Squeezing}. The
imprints in the Wigner function are demonstrated in Section \ref{Sect-Wigner}%
. We clarify the mechanisms in Section \ref{Sect-Mechanisms}. An overview of
the phase transitions is given in Section \ref{Sect-Overall-View} and
finally Section \ref{Sect-Conclusions} is devoted to conclusions and
discussions.

\section{Model and Symmetry}

\label{Sect-Model}

The QRM and the JCM differ in the CRT which can be continuously tuned via
the anisotropic QRM with the following Hamiltonian
\begin{equation}
H=\omega a^{\dagger }a+\frac{\Omega }{2}\sigma _{x}+g\left[ \left(
\widetilde{\sigma }_{-}a^{\dagger }+\widetilde{\sigma }_{+}a\right) +\lambda
\left( \widetilde{\sigma }_{+}a^{\dagger }+\widetilde{\sigma }_{-}a\right) %
\right]
\end{equation}%
where $a^{\dagger }(a)$ creates (annihilates) a bosonic mode with frequency $%
\omega $ and $\sigma _{x,y,z}$ is the Pauli matrix. The coupling with
strength $g$ includes the rotating-wave term and afore-mentioned
counter-rotating term (CRT), the latter is controlled by the anisotropy $%
\lambda $. One can retrieve the QRM and the JCM by setting $\lambda =1$ and $%
\lambda =0$ respectively. Here the unconventional definition of spin raising
and lowering operators $\widetilde{\sigma }^{\pm }=(\sigma _{z}\mp i\sigma
_{y})/2$ is due to the adoption of the spin notation as in ref.\cite%
{Irish2014}, in which $\sigma _{z}=\pm $ conveniently represents the two flux
states in the flux-qubit circuit system\cite{flux-qubit-Mooij-1999} which is
a mostly used platform for realization of ultra-strong couplings. \cite%
{Diaz2019RevModPhy,Wallraff2004,Gunter2009,Niemczyk2010,Peropadre2010,FornDiaz2017, Forn-Diaz2010,Scalari2012,Xiang2013,Yoshihara2017NatPhys,Kockum2017}
The conventional form of the QRM and the JCM can be recovered by replacement
\{$\sigma _{x},\sigma _{y},\sigma _{z}$\} $\rightarrow $ \{$\sigma
_{z},-\sigma _{y},\sigma _{x}$\} via a spin rotation around the axis $\vec{x%
}+\vec{z}$, with the $\Omega $ term denoting the atomic level splitting in
cavity systems. The model has the parity symmetry, $[\hat{P},H]=0$ with $%
\hat{P}=\sigma _{x}(-1)^{a^{\dagger }a}$, at any anisotropy.

To facilitate our analysis we map the Hamiltonian to the effective spatial
space
\begin{equation}
H=\frac{\omega }{2}\hat{p}^{2}+v_{\sigma _{z}}+[\frac{\Omega }{2}-g_{y}i%
\sqrt{2}\hat{p}]\sigma ^{+}+[\frac{\Omega }{2}+g_{y}i\sqrt{2}\hat{p}]\sigma
^{-}  \label{Hx}
\end{equation}%
by the quadrature representation, $a^{\dagger }=(\hat{x}-i\hat{p})/\sqrt{2},$
$a=(\hat{x}+i\hat{p})/\sqrt{2}$, with momentum $\hat{p}=-i\frac{\partial }{%
\partial x}$, as well as spin raising and lowering on the $\sigma _{z}=\pm $
basis, $\sigma _{x}=\sigma ^{+}+\sigma ^{-}$, $\sigma _{y}=-i(\sigma
_{+}-\sigma _{-})$. Let us define the dimensionless anisotropic coupling
strengths $g_{y,z}^{\prime }=\sqrt{2}g_{y,z}/\omega $ for $g_{y}=\frac{%
\left( 1-\lambda \right) }{2}g$ and $g_{z}=\frac{\left( 1+\lambda \right) }{2%
}g$. In such a formalism, the coupling effectively drives a displacement by $%
g_{z}^{\prime }$ in the harmonic potentials $v_{\sigma _{z}}=\omega \left(
x+g_{z}^{\prime }\sigma _{z}\right) ^{2}/2+\varepsilon _{0}^{z}$ in opposite
directions for the two spin components, up to a constant $\varepsilon
_{0}^{z}=-\frac{1}{2}[g_{z}^{\prime 2}+1]\omega $. Then, rather than the
afore-mentioned atomic splitting, the $\Omega $ term now plays the role of
spin flipping in the spin $\sigma _{z}$ space and the role of tunneling in the
effective spatial space.\cite{Ying2015,Irish2014} The $g_{y}$ term actually
resembles the Rashba spin-orbit coupling.

The Hamiltonian can be re-arranged to be an explicitly $x$-$p$ dual form
\begin{eqnarray}
H_{x} &=&\frac{\omega }{2}[(-i\frac{\partial }{\partial x}+g_{y}^{\prime
}\sigma _{y})^{2}+\left( x+g_{z}^{\prime }\sigma _{z}\right) ^{2}]+\frac{%
\Omega }{2}\sigma _{x}+\varepsilon _{0},  \label{H2-x} \\
H_{p} &=&\frac{\omega }{2}[(-i\frac{\partial }{\partial p}-g_{z}^{\prime
}\sigma _{z})^{2}+\left( p+g_{y}^{\prime }\sigma _{y}\right) ^{2}]+\frac{%
\Omega }{2}\sigma _{x}+\varepsilon _{0},   \label{H2-p}
\end{eqnarray}
where $\varepsilon _{0}=-\omega (1+g_{z}^{\prime 2}+g_{y}^{\prime 2})/2$ and
we have used $\hat{x}=i\frac{\partial }{\partial p}$ which fits $[\hat{x},\hat{p}]=i$. From (\ref{H2-x}) and (\ref{H2-p}) one can see directly that the
positive-$\lambda $ and the negative-$\lambda $ regime are symmetric under
the spin rotation and transform to momentum space
\begin{equation}
\left\{ \sigma _{x},\sigma _{y},\sigma _{z}\right\} \rightarrow \left\{
\sigma _{x},-\sigma _{z},\sigma _{y}\right\} ,\quad x\rightarrow p,\quad
\lambda \rightarrow -\lambda .  \label{duality-exchange}
\end{equation}%
Note that in polaron picture the quantum phase transition at $g_{c}^{\lambda
}=\frac{2}{1+\left\vert \lambda \right\vert }g_{\mathrm{s}}$, with
$g_{\mathrm{s}}=\sqrt{\omega \Omega }/2$, essentially is a wave packet splitting
from a Gaussian-like wave packet into two wave packets in the potential
separation.\cite{Ying2015,Ying-2021-AQT} In the positive-$\lambda $ regime,
comparing (\ref{Hx}) with its dual form indicates that $g_{z}^{\prime }$
with a larger amplitude can generate a larger potential separation in $x$
space than $g_{y}^{\prime }$ in $p$ space to bring the quantum phase
transition, while the $g_{y}^{\prime }$ term in $x$ space and the $%
g_{z}^{\prime }$ term in $p$ space play little role for the quantum phase
transition due to self-cancelation as the derivative of the Gaussian-like wave
packet before the transition is an odd function.\cite{Ying-2021-AQT} Thus
the positive-$\lambda $ regime is $x$-type in the sense that $\langle \hat{x}%
^{2}\rangle $ is more dominant than $\langle \hat{p}^{2}\rangle $, and vice
versa, the negative-$\lambda $ regime is $p$-type.\cite%
{LiuM2017PRL,Ying-2021-AQT} Hereafter we discuss in the $x$ space for $\lambda
>0 $, while the result is symmetrically available in the $p$ space for $\lambda
>0$.

%%%%%%%%%%%%%%%%%%%%%%%%%%%%%%%%%%%%%%%%%%%%%%%%%%%%%%%%%%%%%%%%%%%%%%%%%%%%%%%%%%%%%%%%%%%%%%%%%%
%\begin{figure*}[tbph]
\begin{figure*}[t]
\centering
\includegraphics[width=2.0\columnwidth]{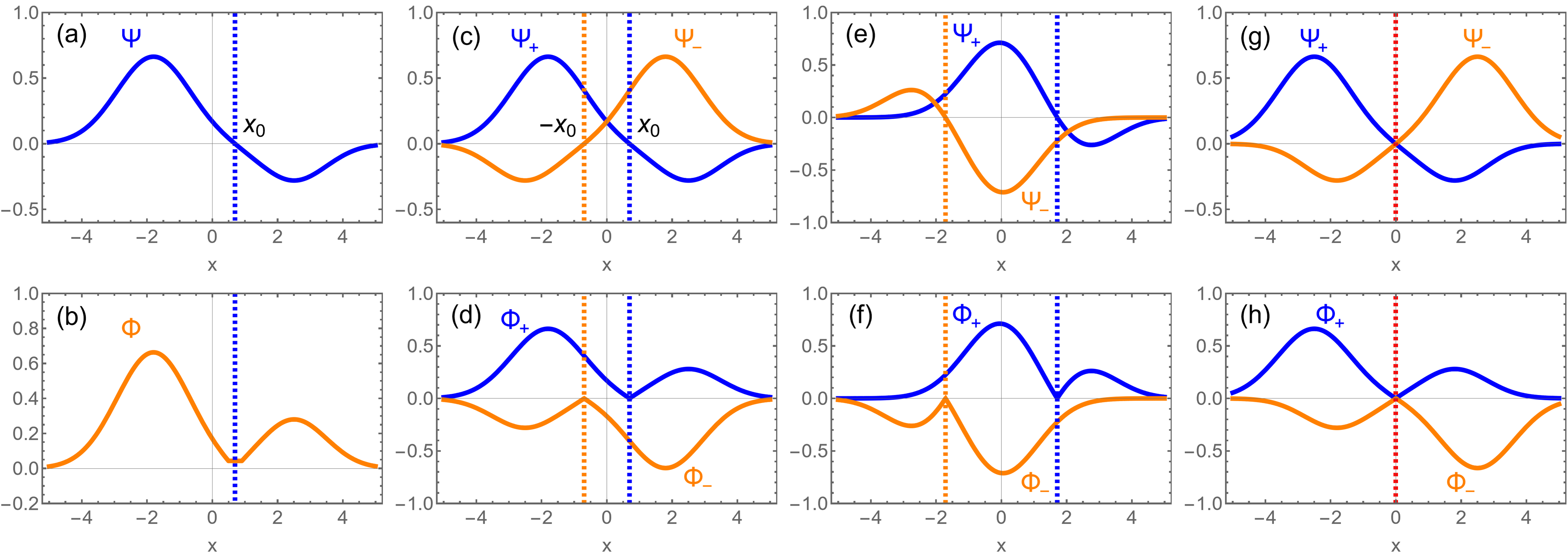}
\caption{\textit{No-node theorem for the QRM.} a) Wave function of a
spinless state $\Psi$ with a node. The dash line marks the node position at $%
x_0$. b) A state $\Phi$ with the node removed by deformation from the absolute
value of $\Psi$ and a round-off around the node. c-h) Spin-up ($+$, blue) and
spin-down ($-$, orange) components of the wave function before ($\Psi_{\pm}$%
) and after ($\Phi_{\pm}$) the deformation, with the nodes located between
the main peaks (c,d), outside the main peaks (e,f) and at the origin (g,h).
The dashed lines in c-h) mark the nodes at $\pm x_0$ and $x_0=0$ in g,h). }
\label{fig-NoNode-theorem}
\end{figure*}
%%%%%%%%%%%%%%%%%%%%%%%%%%%%%%%%%%%%%%%%%%%%%%%%%%%%%%%%%%%%%%%%%%%%%%%%%%%%%%%%%%%%%%%%%%%%%%%%%%

\section{No-Node Theorem for Quantum Rabi Model}
\label{Sect-NoNode-Rabi}

Let us start with the no-node theorem which is general for one-dimensional
(1D) spinless systems.\cite{Ref-No-node-theorem} A node is a zero point in the wave function.
The no-node theorem tells us that for a 1D system with bound states in a
local potential $V(x)$ the ground state has no nodes. For the sake of
further extension to the QRM, let us follow the conventional proof of the
theorem by a deformation from a wave function $\Psi (x)$ with a node to its
absolute value $\Phi (x)$ with the node eliminated by a round-off within a
small interval $\epsilon $. As sketched in \textbf{Figure \ref%
{fig-NoNode-theorem}}a,b, for an order estimation of the energy change the
round-off can be approximated by a constant cut at the interval, i.e., by
assuming $\Psi (x)\approx -k(x-x_{0})$ around the node $x_{0}$ one sets $%
\Phi (x)=Nk\epsilon $ for $\left\vert x-x_{0}\right\vert \leqslant \epsilon $%
; otherwise $\Phi (x)=N\left\vert \Psi (x)\right\vert $ for $\left\vert
x-x_{0}\right\vert >\epsilon $, where $N=1/\sqrt{1+4k^{2}\epsilon ^{2}/3}$
is the renormalization factor. Here the gradient $k$ around the node is
nonzero as the quantum mechanics excludes simultaneous vanishing of the wave
function and its derivation at a same point, otherwise the wave function
would have been suppressed completely. Note $\int\limits_{-\infty }^{\infty }\Psi
^{\ast }\partial _{x}^{2}\Psi dx=-\int\limits_{-\infty }^{\infty }\left\vert
\partial _{x}\Psi \right\vert ^{2}dx$ for bound states, it can be readily
shown that this deformation leads to a decrease of the kinetic energy by an
order $\epsilon $ while both the renormalization and the potential energy
yield a variation of subdominant order $\epsilon ^{3}$.\cite%
{Ref-No-node-theorem} Thus, a nodeless state is more favorable for the
ground state.

Note that the no-node theorem applies for spinless particles or systems
without spin interactions, now we extend the theorem to the QRM ($\lambda =1$%
) which involves communications of the spin components via both the tunneling $\Omega $ term
and the coupling $g$ term. The formalism (\ref{Hx}) facilitates our analysis
since the coupling $g$ term effectively gives rise to the harmonic
potentials $v_{\sigma _{z}}(x)$ displaced in opposite directions for the two
spin components. In the absence of the tunneling both the two spin
components are effectively 1D systems which obey the no-node theorem. In the
presence of the tunneling we can apply the similar wave-function deformation
but now for both the spin components, $\left\vert \Psi \right\rangle =\left(
\Psi _{+}\left\vert +\right\rangle +\Psi _{-}\left\vert -\right\rangle
\right) /\sqrt{2}$, as in Figure \ref{fig-NoNode-theorem}c-h. Note that the
system has the parity symmetry $\Psi _{-}(x)=P\Psi _{+}(-x), $ where $P=\pm 1$,
which involves the space inversion $x\rightarrow -x$ besides the spin
reversion, the tunneling energy is then
\begin{equation}
E_{\Omega }=P\frac{\Omega }{2}\int\limits_{-\infty }^{\infty }\Psi
_{+}(x)\Psi _{+}(-x)dx,
\end{equation}
where $\Psi _{+}$ is chosen to be a real function as the ground state is non-degenerate.

%%%%%%%%%%%%%%%%%%%%%%%%%%%%%%%%%%%%%%%%%%%%%%%%%%%%%%%%%%%%%%%%%%%%%%%%%%%%%%%%%%%%%%%%%%%%%%%%%%
\begin{figure*}[t]
\centering
\includegraphics[width=1.5\columnwidth]{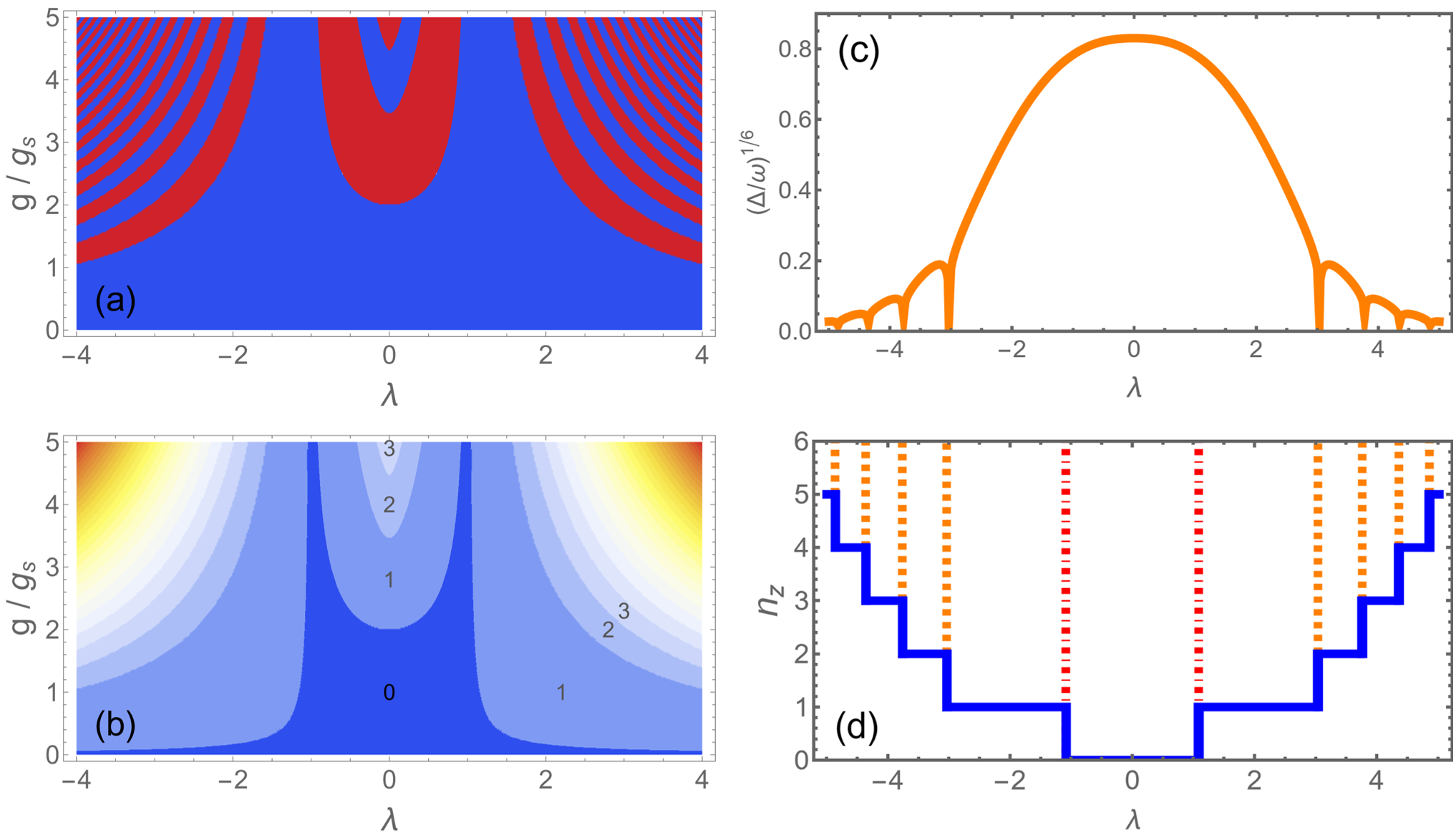}
\caption{\textit{Topological transitions with and without gap closing.} a)
Ground-state (GS) phase diagram of parity in $\protect\lambda$-$g$ plane b)
GS phase diagram of the node number $n_Z$ in $\protect\lambda$-$g$ plane, the
numbers mark some of $n_Z$ which increases by one for each stripe from blue
to red regimes. c) The first excitation gap $\Delta$ versus $\protect\lambda$
at $g=1.5g_{\mathrm{s}}$. d) $n_Z$ (blue solid line) versus $\protect\lambda$
at $g=1.5g_{\mathrm{s}}$, the orange dashed lines mark the transitions with
gap closing corresponding to c) while the red dot-dashed lines remind the
transitions without gap closing. Here, $\protect\omega=0.5\Omega$.}
\label{fig-gap-P-nz}
\end{figure*}
%%%%%%%%%%%%%%%%%%%%%%%%%%%%%%%%%%%%%%%%%%%%%%%%%%%%%%%%%%%%%%%%%%%%%%%%%%%%%%%%%%%%%%%%%%%%%%%%%%

Unlike the spinless systems in the conventional no-node theorem, here in the space inversion the node position $x_{0}$ becomes relevant and there
are three possible situations: (i) In the first situation, as in Figure \ref%
{fig-NoNode-theorem}c, the nodes in the two spin components are located
between the origin and the main peaks of $\Psi _{\pm }(x),$ in such a case a
positive parity would be favorable to gain more negative tunneling energy
around the peaks. After the deformation, as compared in Figure \ref%
{fig-NoNode-theorem}c,d, $\Phi (x)$ has the same the tunneling energy
contribution from $\left\vert x\right\vert >\left\vert x_{0}\right\vert $ as
$\Psi (x)$ but gains more negative tunneling energy from $\left\vert
x\right\vert <\left\vert x_{0}\right\vert $ due to the opposite signs of $%
\Phi _{+}$ and $\Phi _{-}$ while $\Psi _{+}$ and $\Psi _{-}$ have the same
sign. Unlike the afore-mentioned small local reduction in the kinetic energy
of order $\epsilon $, the decrease in the tunneling energy is more global
and finite thus being dominant. (ii) In the second situation, as in Figure %
\ref{fig-NoNode-theorem}e, the nodes appear outside the main peak region so
that $\Psi $ has a negative parity with opposite signs of the main peaks. In
this case, the tunneling energy after the deformation as in Figure \ref%
{fig-NoNode-theorem}f remains the same for $\left\vert x\right\vert
<\left\vert x_{0}\right\vert $ but has a finite decrease for $\left\vert
x\right\vert >\left\vert x_{0}\right\vert .$ (iii) In the third situation,
the node is located at the origin, which needs more delicate consideration
as the energy variation order now depends on the round-off. In this case $%
P=1 $ for $\Psi $ and $P=-1$ for $\Phi $. Without the round-off, the
tunneling energy before ($E_{\Omega }^{\Psi }$) and after ($E_{\Omega
}^{\Phi }$) the deformation are equal. Taking the round-off into account we
have a decrease of tunneling energy by order $\epsilon ^{3}$ (see derivation
in Appendix \ref{Appendix-no-node}):%
\begin{equation}
E_{\Omega }^{\Phi }-E_{\Omega }^{\Psi }=-\left( \frac{4}{3}k^{2}\Delta
_{\rho }\Omega \right) \ \epsilon ^{3},
\end{equation}%
where $\Delta _{\rho }=\left\vert \sum_{n=0}^{\infty }\left\vert
C_{n}\right\vert ^{2}\right\vert -\left\vert \sum_{n=0}^{\infty
}(-1)^{n}\left\vert C_{n}\right\vert ^{2}\right\vert >0$ and $C_{n}$ is the
expansion coefficient on the basis of quantum harmonic oscillator $\Psi
_{+}(x)=\sum_{n}C_{n}\phi _{n}$.

Thus the deformation always leads to an energy decrease from the tunneling energy, no
matter finite or small, in addition to the energy reduction in the no-node
theorem before the tunneling is added. Therefore, we can conclude that the
ground state of the QRM\ has no nodes either.

\section{States with Nodes and Conventional Topological Transitions with Gap
Closing}

\label{Sect-Top-with-gap-closing}

Note that the QRM has an isotropic coupling with equal weights ($\lambda =1$%
) of the rotating-wave term and counter-rotating term, which together just
right play the role of displacing the potentials by $g^{\prime }$. In the
presence of the coupling anisotropy an extra Rashba spin-orbit coupling
arises as the $g_{y}$ term in the Hamiltonian (\ref{Hx}). This $g_{y}$ term
couples the spin components with a strength depending on the gradient of the
wave function, thus inducing an effect very different from the direct spin
flipping by the $\Omega $ term. Such a coupling breaks the limitation of the
no-node theorem and brings about a series of phase transitions and a rich
phase diagram.

As extracted by exact diagonalization,\cite{Ying2020-nonlinear-bias} \textbf{%
Figure \ref{fig-gap-P-nz}}a shows the phase diagram of the ground-state
parity, by the example at $\omega =0.5\Omega $, where red and blue colors represent positive
and negative parities respectively. A transition occurs each time when the
parity is reversed. As illustrated by Figure \ref{fig-gap-P-nz}c, the parity
reversal is always accompanied with a gap closing as the lowest two energy
levels with different parities are crossing each other. Such transitions
occur in the regime $\left\vert \lambda \right\vert \leqslant 1$ which
connects the QRM ($\lambda =1$) and the JCM ($\lambda =0$) and were
found to be topological phase transitions.\cite{Ying-2021-AQT} The
transitions change the node number of the ground state, $n_{Z}$, which
characterizes the topological structure of the ground state, since by
keeping a fixed number of $n_{Z}$ one cannot go to another $n_{Z}$ state by
continuous shape deformation of the wave function, just as one cannot change
a doughnut (torus) into a sphere by a continuous deformation.
Note that generally speaking there is no universal topological invariant for all systems even in condensed matter.\cite{TopCriterion} The
mostly used Chern number based on many-body Brillouin zone structure \cite{TopCriterion,TopNori,Top-Guan} is not either applicable for few-body systems. However,
the node number provides a topological quantum number for the anisotropic QRM that is invariant within a topological phase and capable of distinguishing different phases of the ground state.  Here we see
that such TPTs in the anisotropic QRM not only occur in $\left\vert \lambda \right\vert \leqslant 1$
regime but also in $\lambda \leqslant 1\ $regime. Indeed, by increasing $%
\lambda $ or $g$ the ground state experiences an infinite series of TPTs,
thus forming a series stripes of phases.

%%%%%%%%%%%%%%%%%%%%%%%%%%%%%%%%%%%%%%%%%%%%%%%%%%%%%%%%%%%%%%%%%%%%%%%%%%%%%%%%%%%%%%%%%%%%%%%%%%
\begin{figure*}[t]
\centering
\includegraphics[width=2\columnwidth]{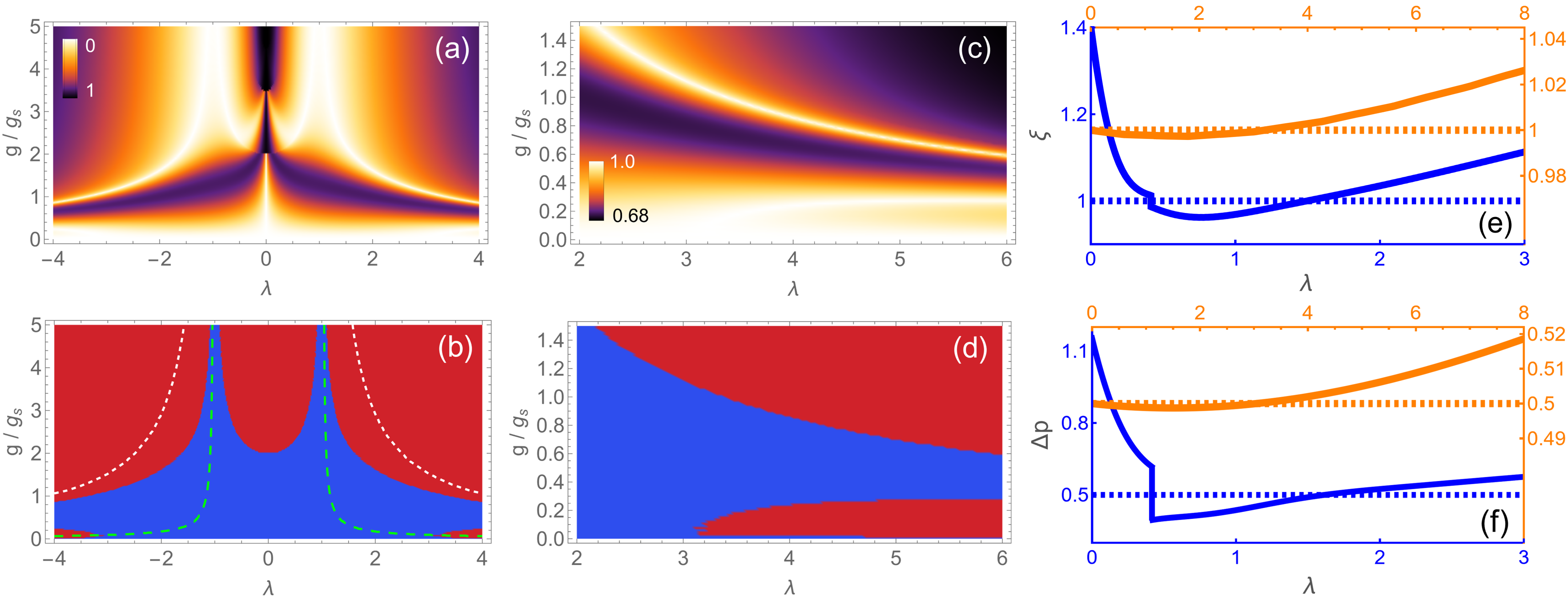}
\caption{\textit{Transitions from amplitude squeezing to phase squeezing.}
a) $|\protect\xi -1|$ in $\lambda$-$g$ plane. b) Sign of $\xi -1$ in $\lambda$-$g$ plane, blue is negative and red is
positive. $\Delta p-1/2$ has a similar phase diagram. c) $|\protect\xi -1|$ in
small-$g$ and large-$\protect\lambda$ regime. The white dotted line is the first
topological transition boundary with gap closing in $\lambda >1$ regime
and the green dashed line is the topological transition boundary without gap
closing. d) Sign of $\protect\xi -1$ in small-$g$ and large-$\protect\lambda$
regime. $\Delta p-1/2$ has a similar phase diagram. e) $\protect\xi$ versus $%
\protect\lambda$ at fixed $g=2.2g_{\mathrm{s}}$ (blue) and $g=0.1g_{\mathrm{s%
}}$ (orange). f) $\Delta p$ versus $\protect\lambda$ at fixed $g=2.2g_{%
\mathrm{s}}$ (blue) and $g=0.1g_{\mathrm{s}}$ (orange). Here, $\protect\omega%
=0.5\Omega$.}
\label{fig-Squeezing}
\end{figure*}
%%%%%%%%%%%%%%%%%%%%%%%%%%%%%%%%%%%%%%%%%%%%%%%%%%%%%%%%%%%%%%%%%%%%%%%%%%%%%%%%%%%%%%%%%%%%%%%%%%

\section{Hidden Unconventional Topological Transitions Without Gap Closing}

\label{Sect-Top-without-gap-closing}

As mentioned in Introduction, in condensed matter the conventional TPTs
occur at gap closing,\ while there are also some unconventional TPTs that
happen without gap closing.\cite{Amaricci-2015-no-gap-closing,Xie-QAH-2021}
We find such an unconventional case also can appear in the anisotropic QRM. A
full phase diagram of the ground-state node number $n_{Z}$ is figured out in
Figure \ref{fig-gap-P-nz}b where the numbers mark $n_{Z}$ which is zero at
QRM lines$\ \left\vert \lambda \right\vert =1$ while it increases by steps
when anisotropy is introduced. For the regime $\left\vert \lambda
\right\vert \leqslant 1$ and the regime with a large $\left\vert \lambda
\right\vert $ the transitions in the node number correspond one by one\ to
those in the\ phase diagram of the parity in Figure \ref{fig-gap-P-nz}b
where the parity changes are accompanied with gap closing. However, as one
can notice, another transition boundary of the node number in Figure \ref%
{fig-gap-P-nz}b emerges around $ \lambda  =1$, with its duality around $ \lambda  =-1$, but has
no match of parity change in Figure \ref{fig-gap-P-nz}a. A clearer
illustration at a fixed coupling $g=1.5g_{\mathrm{s}}$ is presented in
Figure \ref{fig-gap-P-nz}d where the orange dashed lines mark the
transitions of $n_{Z}$ in large $\lambda $ that correspond to the gap
closings in Figure \ref{fig-gap-P-nz}b, whereas the transitions labeled by
the red dot-dashed lines around $\left\vert \lambda \right\vert =1$ have no
gap closing. We find that the conventional transitions add nodes around the
origin, as from Figure \ref{fig-NoNode-theorem}d,h to\ Figure \ref%
{fig-NoNode-theorem}c,g which change the parity, while the unconventional
transition introduces a node from the infinity as from Figure \ref%
{fig-NoNode-theorem}f to\ Figure \ref{fig-NoNode-theorem}e which does not
change the parity. We shall clarify the underlying mechanisms later on in
Section \ref{Sect-Mechanisms}.

Unlike the TPTs with parity change and gap closing which are also
accompanied with first-order transitions in some physical properties,\cite%
{Ying-2021-AQT} here the unconventional TPT has neither parity change nor
gap closing to induce any abrupt changes of physical properties. The
unconventional transition would be of infinite order as the node is coming
from the infinity. The transition seems to completely lie in the topological feature
of the ground-state wave function with the node number being invariant
within a topological phase but changing across the phase boundary. Thus, the unconventional transition is hidden in
the sense that it would be not easy to find unless one looks closely into the
topological structure of the ground state. Nevertheless, although right at
the unconventional transition there is no obvious change of physical
properties apart from the change of the topological invariant, lagging
behind the unconventional transition we can find a squeezing-type transition
which can be regarded as a hysteresis sign of the unconventional transition,
as addressed in the following.

\section{Transitions from Amplitude Squeezing to Phase Squeezing
in Gapped Phase}
\label{Sect-Squeezing}

After the afore-mentioned unconventional TPT in increasing the anisotropy
from the QRM line, we also find a transition from amplitude squeezing (AS)
to phase squeezing (PS) in the gapped phase. Actually, rather than polarons
with coherent-state expansion for each wave packet,\cite{Bera2014Polaron}
one wave packet in decomposing the wave function can be more compactly
represented by a frequency-renormalized polaron depicted by\cite{Ying2015}
\begin{equation}
\varphi (\xi ,\zeta )=e^{-\xi (x-\zeta g^{\prime })^{2}/2}\left( \xi /\pi
\right) ^{1/4}  \label{Gaussian-polaron}
\end{equation}%
with the displacement renormalization $\zeta $ and the frequency
renormalization $\xi $.  The squeezing \cite{Ref-Squeezing} is
reflected by the frequency renormalization $\xi $, with $\xi <1$ indicating
an AS with a wave-packet extension in the $x$ space while $\xi >1$
characterizing a PS with a wave-packet extension in the $p$ space. We can
extract $\xi $ for the main peak of the wave function by the ratio
\begin{equation}
\xi =\left( r_{G}/r_{\psi }\right) ^{2}
\end{equation}%
of the Gaussian half-peak radius $r_{G}=\sqrt{2\ln 2}$ and the wave-function
half-peak radius $r_{\psi }$. To reduce the influence of the possible
secondary wave packet, $r_{\psi }$ is extracted from the distance between
the peak position and the half-peak position on the side farther from the
origin. We show the absolute deviation of $\xi $ from $1$, i.e. $\left\vert
\xi -1\right\vert ,$ in \textbf{Figure \ref{fig-Squeezing}}a where the
bright lines indicate the AS/PS transition with $\xi =1$. The transition
boundaries can be more clearly seen by the sign of the $\xi $ deviation from $%
\xi =1$ in Figure \ref{fig-Squeezing}b where the blue and red regions have
an AS ($\xi -1<0$) and a PS ($\xi -1>0$) respectively. Another quantity that
can reflect the squeezing is the fluctuation or the variance of the
momentum, $\Delta p=\langle \hat{p}^{2}\rangle -\langle \hat{p}\rangle ^{2}$%
, with the AS and the PS indicated by $\Delta p<1/2$ and $\Delta p>1/2$. The
phase diagram of $\Delta p-1/2$ is similar to that of $\xi -1$ in Figure \ref%
{fig-Squeezing}b, except for some small boundary discrepancy due to that $%
\xi $ is extracted from the main peak of wave packet while $\Delta p$ is the
expectation over the total wave function.

In $\lambda <1$ regime, the AS/PS boundary coincides with the first boundary of the conventional TPTs.
However, in $\lambda >1$ regime, the AS/PS boundary in Figure \ref{fig-Squeezing}b appears between the first
conventional TPT in $\lambda >1$ regime (dotted line) and the unconventional topological transition (dashed
line). The AS/PS transition occurs before the dotted line where the parity
changes and the gap closes for the first time in entering the $\lambda >1$ regime as in Figure \ref{fig-gap-P-nz}a,c, thus also without gap closing as
the unconventional TPT at the dashed line. In other words, this AS/PS transition is lagging
behind the dashed line where the first node enters the ground state when one
leaves the nodeless QRM line at $\lambda =1$. In Section \ref%
{Sect-Mechanisms} we will see this AS/PS transition is coming retarded due
to awaiting enough strengthening of the nodal state after the unconventional
TPT. Apart from the main AS/PS transition boundary in large-$g$ regime, a
peculiar AS/PS transition is also hidden in small-$g$ and large-$\lambda $
regime in Figure \ref{fig-Squeezing}b and a zoom-in view is provided in
Figure \ref{fig-Squeezing}c,d. Examples of $\xi $ and $\Delta p$ at fixed
couplings are illustrated in Figure \ref{fig-Squeezing}e,f which show the
AS/PS transitions both in large-$g$ regime (blue) and small-$g$ regime
(orange), as indicated by the dotted lines.

\section{Imprints in the Wigner Function}
\label{Sect-Wigner}

%%%%%%%%%%%%%%%%%%%%%%%%%%%%%%%%%%%%%%%%%%%%%%%%%%%%%%%%%%%%%%%%%%%%%%%%%%%%%%%%%%%%%%%%%%%%%%%%%%
\begin{figure*}[tbph]
\centering
\includegraphics[width=2\columnwidth]{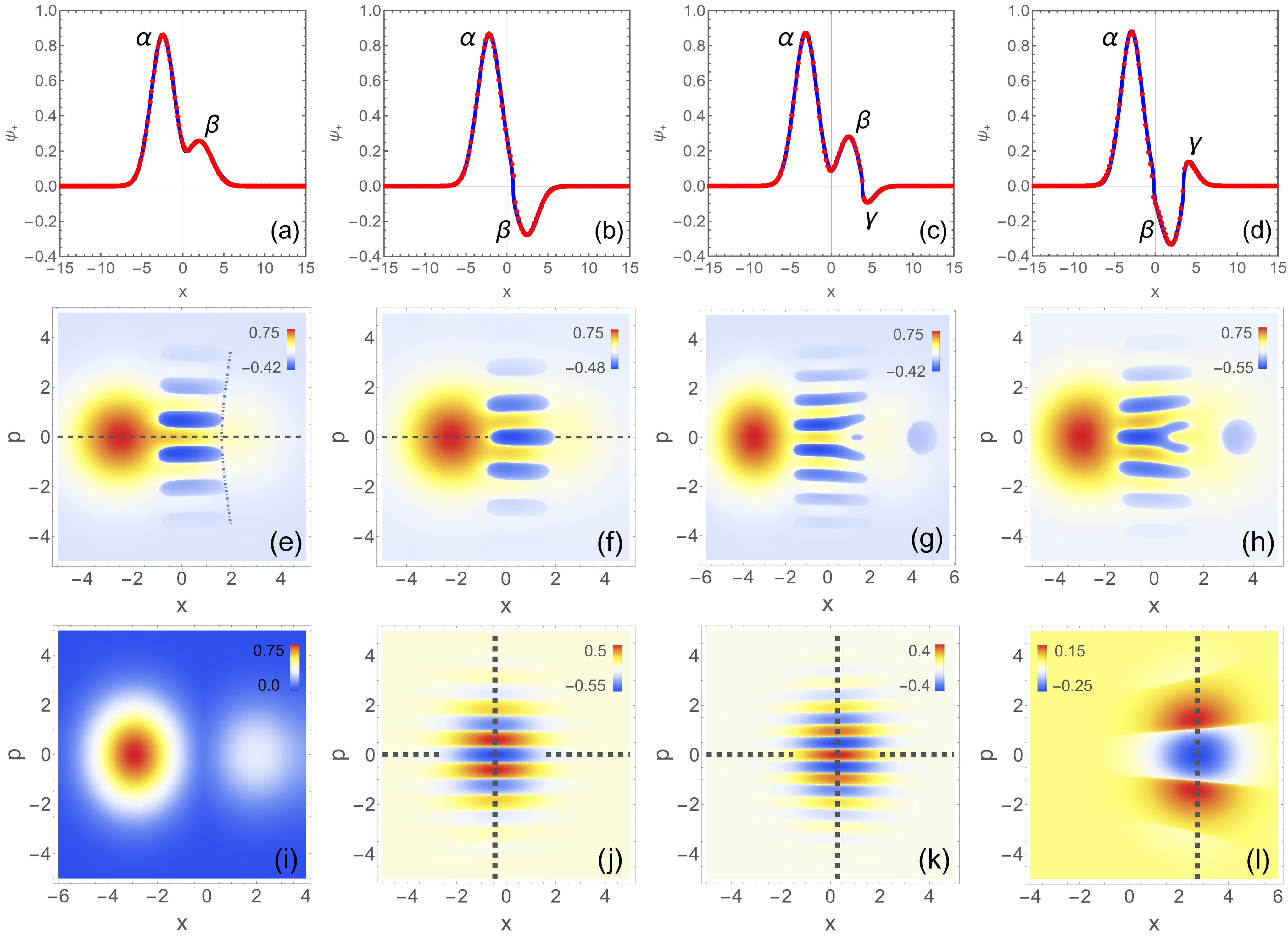}
\caption{\textit{Imprints of nodal status in the Wigner function.} a-d)
Spin-up component of GS Wave function $\protect\psi _+$ at $\{\protect\lambda%
, g/g_{\mathrm{s}} \} = \{1.0,2.5\}$ ($n_Z=0,P=-1$), $\{0.5,3.0\}$ ($%
n_Z=1,P=1$), $\{2.5,1.8\}$ ($n_Z=1,P=-1$), and $\{4.0,1.2\}$ ($n_Z=2,P=1$).
Here the red dots are results of exact diagonalization (ED) while the solid lines
are plotted by polaron picture (see parameters in Appendix \protect\ref%
{Appendix-polaron-parameters}).\protect\cite{Ying2015} e-h) The Wigner
function $W(x,p)$ corresponding to a-d) by numeric integrals based on the ED
wave function. The dashed lines in e),f) remind central or not for the negative peak position (blue) of
the fringe. The dot-dashed line in e) is a guide to the eye for the curved
fringe alignments. i-l) $W(x,p)$ in h) decomposed by three polarons into i)
inner-polaron terms $\sum _{i=\protect\alpha,\protect\beta,\protect\gamma}
W_{ii}(x,p)$ and inter-polaron terms j) $W_{\protect\alpha \protect\beta}$,
k) $W_{\protect\alpha \protect\gamma}$, and l) $W_{\protect\beta \protect%
\gamma}$. The horizontal and vertical dashed lines remind the central-peak
signs and the positions $x_{ij}$ of the decomposed fringes. Here, $\protect\omega%
=0.5\Omega$ and the wave-function amplitude is amplified by $|\protect\psi %
_+|^{1/2}$ to show the node more clearly and the Winger-function amplitude
is amplified by $|W|^{1/4}$ to enhance the color contrast.}
\label{fig-Wigner}
\end{figure*}
%%%%%%%%%%%%%%%%%%%%%%%%%%%%%%%%%%%%%%%%%%%%%%%%%%%%%%%%%%%%%%%%%%%%%%%%%%%%%%%%%%%%%%%%%%%%%%%%%%

Both the squeezing and the nodal status may leave imprints in the Wigner
function which is defined in the phase space and enables the visualization
of information of the momentum:\cite{Winger1932,WingerReview2018}
\begin{equation}
W\left( x,p\right) =\frac{1}{2\pi }\int\limits_{-\infty }^{\infty
}e^{ipy}\psi ^{\dagger }(x+\frac{y}{2})\psi (x-\frac{y}{2})dy,
\end{equation}%
where $\left\vert \psi \right\rangle =\left( \psi _{+}\left\vert
+\right\rangle +\psi _{-}\left\vert -\right\rangle \right) /\sqrt{2}$ and we
have set $\hbar =1$. We show four typical examples of the ground-state wave
function in \textbf{Figure \ref{fig-Wigner}}a-d and the corresponding Wigner
function in Figure \ref{fig-Wigner}e-h: (i) Figure \ref{fig-Wigner}a
represents a nodeless state at $\lambda =1$, the Wigner function of which
has negative interference fringes (blue) away from the central line at $p=0$
as shown in Figure \ref{fig-Wigner}e. (ii) Figure \ref{fig-Wigner}b
illustrates a state with one node. In contrast to Figure \ref{fig-Wigner}e
the Wigner function in Figure \ref{fig-Wigner}f has a central piece of the
negative interference fringes as indicated by the dashed lines. This central
negative interference fringe (CNIGF) turns out to be a sign of the node.
Note that the node in this case is located around the origin $x=0$. (iii) We show
another nodeful case in Figure \ref{fig-Wigner}c where the node is not
around the origin but appears beyond the secondary peak, thus we have two
positive peaks ($\alpha ,\beta $) and a negative peak ($\gamma $) with the
node between $\beta $ and $\gamma $. The Wigner function in Figure \ref%
{fig-Wigner}g has two regimes of interference fringes: one around $x=-0.5$
has no CNIGF and the other around $x=1.5$ has a CNIGF, later on we will see
the former comes from the $\alpha $-$\beta $ interference while the latter
originates from the $\alpha $-$\gamma $ interference. Although this CNIGF is
small due to the $\alpha $-$\gamma $ distance and the small weight of $%
\gamma $, a large negative central spot emerges around $x=4.5$ which stems
from the $\beta $-$\gamma $ interference. (iv) A state with one node around
the origin and another node beyond the secondary peak is presented in Figure %
\ref{fig-Wigner}d. The Wigner function in Figure \ref{fig-Wigner}h now has a
fringe regime with a CNIGF around $x=-0.5,$ and a fringe regime without
CNIGF around $x=1.5$, and a central negative interference spot around $x=3.5$%
.

To see the origins of the interference fringes and spots we can decompose
the wave function into $n_{p}$ number of frequency-renormalized polarons
with weight $w_{i}$ \cite{Ying2015,CongLei2017}%
\begin{equation}
\psi _{+}=\sum_{i=1}^{n_{p}}w_{i}\varphi (\xi _{i},\zeta _{i}),
\label{WF-polarons}
\end{equation}%
which can reproduce well the wave function from the exact diagonalization (ED),
as compared in Figure \ref{fig-Wigner}a-d where the blue solid lines are
results in polaron picture which are in good agreements with the\ ED (red
dots). With the polaron decomposition (\ref{WF-polarons}), we obtain an
explicit expression of the Wigner function with different contributions from the inner-polaron part $%
W_{ii}$ and the inter-polaron part $W_{ij}$:
\begin{equation}
W\left( x,p\right) =\sum_{i=1}^{n_{p}}W_{ii}\left( x,p\right)
+2\sum_{i<j}^{n_{p}}W_{ij}\left( x,p\right) ,
\end{equation}%
where%
\begin{eqnarray}
W_{ii} &=&\frac{w_{i}^{2}}{\pi N_{p}^{2}}e^{-\frac{p^{2}}{\xi _{i}}-\xi
_{i}\left( x-x_{i}\right) ^{2}}, \\
W_{ij} &=&\frac{w_{i}w_{j}}{\pi N_{p}^{2}}\left( \frac{\xi _{i}\xi _{i}}{\xi
_{ij}^{2}}\right) ^{1/4}e^{-\frac{p^{2}}{\xi _{ij}}-\frac{\xi _{i}\xi _{i}}{%
\xi _{ij}}\left( x-x_{ij}\right) ^{2}}f_{ij}, \\
f_{ij} &=&\cos [p\frac{(x-x_{i})\xi _{i}-(x-x_{j})\xi _{j}}{\xi _{ij}}],
\end{eqnarray}%
and $x_{i}=\zeta _{i}g^{\prime }$ is the polaron position determined by the
Gaussian center while $x_{ij}=(x_{i}+x_{j})/2$ and $\xi _{ij}=(\xi _{i}+\xi
_{j})/2$ are averaged position and frequency renormalization for polaron $i$
and $j$. The normalization factor is decided by $N_{p}=1/\sqrt{%
\sum_{i,j}S_{ij}}$ with $S_{ij}=\left( \xi _{i}\xi _{i}/\xi _{ij}^{2}\right)
^{1/4}\exp [-\frac{\xi _{i}\xi _{i}}{\xi _{ij}}\left( \frac{x_{i}-x_{j}}{2}%
\right) ^{2}]$. Figure \ref{fig-Wigner}i-l are the decomposed inner-polaron
part $W_{ii}$ and inter-polaron part $W_{ij}$ which together well reproduce the
total Wigner function by ED in Figure \ref{fig-Wigner}i-l. We see that
inner-polaron part $W_{ii}$ is round in the phase space if there is no
squeezing $\xi _{i}=1$, otherwise it becomes oval along $x$ direction for
amplitude squeezing with $\xi _{i}<1$ or along $p$ direction for phase
squeezing with $\xi _{i}>1$, as illustrated by Figure \ref{fig-Wigner}i
which has a phase squeezing. The interference fringes are coming from $%
f_{ij} $ in the inter-polaron part which oscillates with an average period%
\begin{equation}
T_{p}=\frac{2\pi }{\left\vert x_{j}-x_{i}\right\vert }  \label{Tp}
\end{equation}%
as the momentum $p$ is increasing, while the fringe position is decided by $%
x_{ij}$. The expression (\ref{Tp}) indicates that a larger polaron distance
leads to denser interference fringes, thus accounting for the period
difference in Figure \ref{fig-Wigner}j-l as the polarons $\beta $ and $%
\gamma $ for $W_{\beta \gamma }$ in Figure \ref{fig-Wigner}l are closer than
those for $W_{\alpha \beta }$ and $W_{\alpha \gamma }$ in Figure \ref%
{fig-Wigner}j,k

It might be worthwhile to mention here that not only inner-polaron part $%
W_{ii}$ and but also inter-polaron part $W_{ij}$ can provide some squeezing
information, with $W_{ij}$ indicating the squeezing difference. In fact,
when the two polarons have different frequencies, the fringe alignment will
not be straight but become curved. The local curvature around $p=0$ can be
extracted to be
\begin{equation}
K=\frac{2\left( \xi _{i}-\xi _{j}\right) }{\left( \xi _{i}+\xi _{j}\right)
\left( x_{j}-x_{i}\right) },
\end{equation}%
which is proportional to the frequency difference. The curving center is on
the side of the polaron with a lower frequency, as illustrated in Figure \ref%
{fig-Wigner}e where the fringe alignment tends to curve around polaron $%
\beta $ as $\xi _{\beta }\approx 0.8$ on the right side is smaller than $\xi
_{\alpha }\approx 0.97$.

Note a nodeless state has polaron weight $w_{i}$ all positive thus $W_{ij}$
is positive around $p=0$, while the node introduces a negative polaron
weight thus turns the positive peak of $W_{ij}$ to be negative around $p=0,$
which accounts for the absence of the CNIGF in Figure \ref{fig-Wigner}e and
the presence of the CNIGF in Figure \ref{fig-Wigner}f. When we have two nodes as
in Figure \ref{fig-Wigner}d,h, the neighboring weight product $\alpha \beta $
and $\beta \gamma $ are negative while the weight of the next neighboring
pair $\alpha \gamma $\ is positive, which produces a CNIGF in Figure \ref%
{fig-Wigner}j,l but not in Figure \ref{fig-Wigner}k as the horizontal lines
remind. Thus, we see both the CNIGF and the negative interference spot at $%
p=0$ are imprints of the nodes. The changes of patterns from e)$\rightarrow$f) and g)$\rightarrow$h)
in Figure \ref{fig-Wigner} are conventional TPTs in $|\lambda| <1$ and $|\lambda| >1$ regimes respectively,
while e)$\rightarrow$g) would indicate the unconventional TPT.

%%%%%%%%%%%%%%%%%%%%%%%%%%%%%%%%%%%%%%%%%%%%%%%%%%%%%%%%%%%%%%%%%%%%%%%%%%%%%%%%%%%%%%%%%%%%%%%%%%
\begin{figure*}[t]
\centering
\includegraphics[width=1.35\columnwidth]{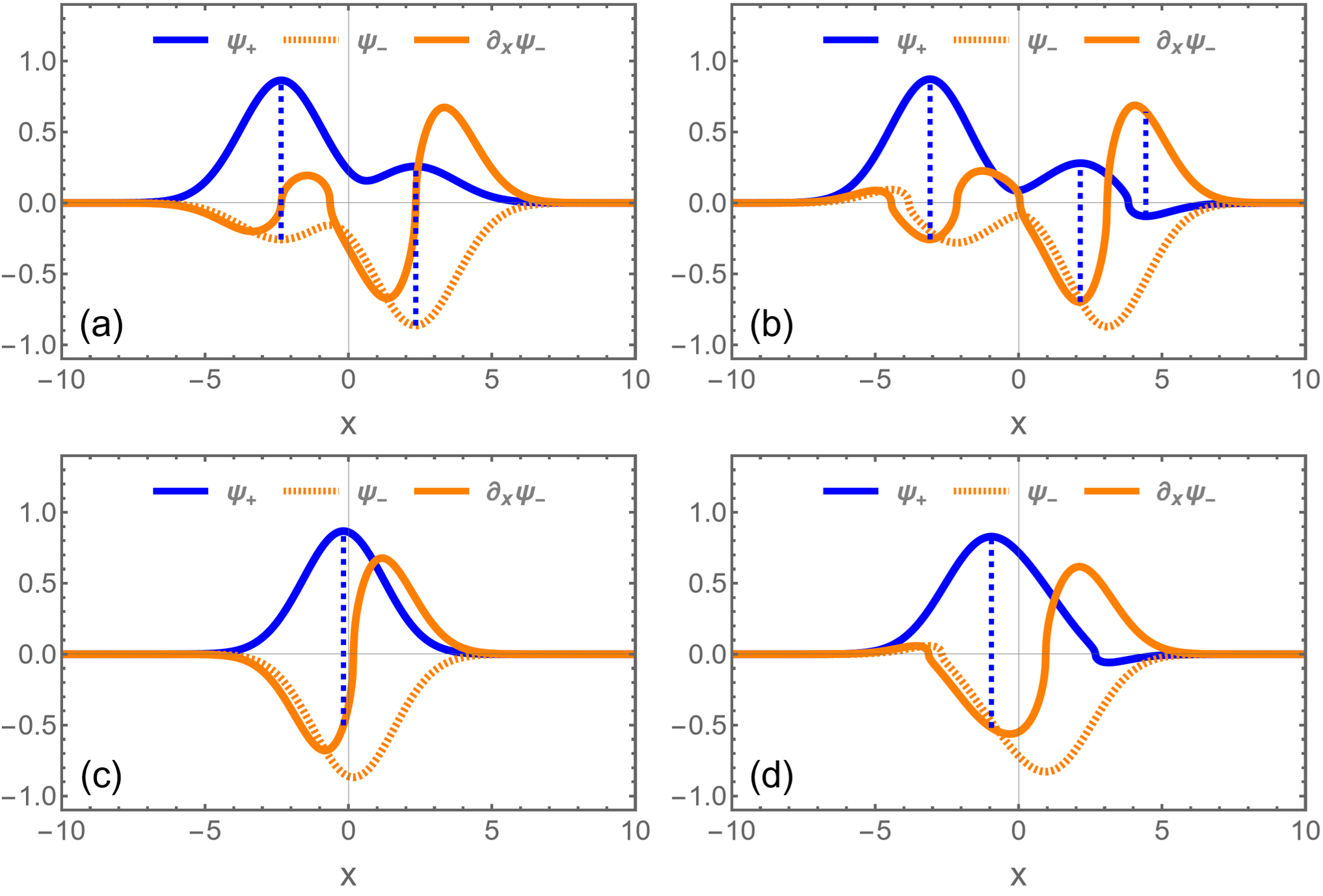}
\caption{\textit{Mechanisms for the topological transitions without gap and
the squeezing transitions.} The GS wave function $\protect\psi _+ (x)$ (blue
solid), $\protect\psi _- (x)$ (orange dotted) and its derivative $\partial
_x \protect\psi _-(x)$ (orange solid) at a) $\protect\lambda =0.7,g=2.8g_{%
\mathrm{s}}$, b) $\protect\lambda =2.5,g=1.8g_{\mathrm{s}}$, c) $\protect%
\lambda =3.0,g=0.8g_{}$, d) $\protect\lambda =5.5,g=0.1g_{\mathrm{s}}$. The
vertical blue dashed lines mark the positions of positive or negative peaks
of $\protect\psi _+ (x)$. Here, $\protect\omega=0.5\Omega$ and to increase
the visibility of the nodes we have plotted the amplitudes by $|\protect\psi %
_\pm|^{1/2}$ and $|\partial _x\protect\psi _-|^{1/2}$. }
\label{fig-Mechanism}
\end{figure*}
%%%%%%%%%%%%%%%%%%%%%%%%%%%%%%%%%%%%%%%%%%%%%%%%%%%%%%%%%%%%%%%%%%%%%%%%%%%%%%%%%%%%%%%%%%%%%%%%%%

\section{Mechanisms}

\label{Sect-Mechanisms}

The topological transition without gap closing and the transitions of the
different types of squeezing actually involve subtle energy competitions.
Clarifying the mechanisms underlying these transitions would gain more
insights for the role of the counter-rotating term in the coupling. In
Section \ref{Sect-NoNode-Rabi} we have seen that competitions of the kinetic
energy, the potential energy and the tunneling energy are not favorable to
introduce a node in the ground state, and the $g_{z}$ term of the
anisotropic coupling in (\ref{Hx}) effectively just contributes to the
potential. So the key to have a nodeful state is in the anisotropic $g_{y}$
term of (\ref{Hx}). We remind here we are focusing on the positive-$\lambda $
regime while the analysis is the same for the negative-$\lambda $ regime by
exchange of $\left\{ g_{y},x\right\} $ and $\left\{ -g_{z},p\right\} $ as
mentioned in (\ref{duality-exchange}). To facilitate the understanding we
write the anisotropic $g_{y}$ term together with the tunneling energy in
the following form
\begin{equation}
E_{\Omega }+E_{g_{y}}=\int\limits_{-\infty }^{\infty }\psi _{+}(x)\left[
\frac{\Omega }{2}\psi _{-}(x)-\sqrt{2}g_{y}\partial _{x}\psi _{-}(x)\right]
dx
\end{equation}%
where the doubling of the contribution from spin exchange has canceled with
the normalization factor $1/\sqrt{2}$. We remind here $g_{y}=\frac{\left(
1-\lambda \right) }{2}g$, the sign of which will make an important
difference.

\subsection{Energy Competitions in Nodeless State}

We first look at a nodeless state in the $\left\vert \lambda \right\vert
\leqslant 1$ regime, as represented by an example in \textbf{Figure \ref%
{fig-Mechanism}}a with $\lambda =0.7$ and $g=2.8g_{\mathrm{s}}$. As the $%
\Omega $ term is more dominant over the small $g_{y}$ with $\lambda $ close
to $1$, the nodeless state has a negative parity to gain more negative
energy in $E_{\Omega }$. In this trend the peaks of $\psi _{+}$ and $\psi
_{-}$ tend to get located at the same positions to get a maximum wave-packet
overlap as indicted by the vertical blue dashed lines in Figure \ref%
{fig-Mechanism}a, while the two-peak structure is due to the interplay of
the tunneling and the potential separation in the two spin components.\cite%
{Ying2015} Note that in this regime $g_{y}$ is positive thus the anisotropic
coupling is actually counter-acting against the $\Omega $ term in a large
region where $\partial _{x}\psi _{-}(x)$ (orange solid line) and $\psi
_{+}(x)$ (blue solid line) have opposite signs. The configuration with same
peak positions of $\psi _{+}$ and $\psi _{-}$ reduces the counteracting
effect of $g_{y}$ to a great degree as it is the zeros of $\partial _{x}\psi _{-}$ that are
meeting the peaks of $\psi _{+}$. On the other hand to gain more negative
energy from $E_{\Omega }$ the wave packets on the right and left sides tend
to get more extended to get right-left overlap, which leads to an amplitude
squeezing.\cite{Ying2015} Such a nodeless status will maintain until the two counteracting energies cancel each other and the
parity reverses at a larger $g_{y}$ at a critical anisotropy $\left\vert
\lambda _{\mathrm{T1}}\right\vert =\sqrt{1-4g_{\mathrm{s}}^{2}/g^{2}}$,\cite%
{Ying-2021-AQT} while different parity represents different quantum states
and parity change means level crossing and gap closing.
The energy cancelation also invalidates the effect of polaron frequency renormalization, thus $\xi$  is returning to $1$ as illustrated by the blue lower line around $\lambda\approx 0.42$ in Figure \ref{fig-Squeezing}e.

\subsection{Mechanism for the Unconventional Topological Transition without Gap Closing}

The $\lambda >1$ regime has a different scenario as now $g_{y}$ reverses the
sign to be negative. In such a situation, the $g_{y}$ term is not
counteracting against the $\Omega $ term but a competing relation arises instead.
When $\lambda $ is large so that the $g_{y}$ term becomes more dominant over
the $\Omega $ term, it is more favorable for the negative peaks of $\partial _{x}\psi
$, rather than those of $\psi _{-}$, to meet the positive $\psi _{+}$ peaks, as shown
by two vertical dashed lines around $x=-3$ and $x=2$ for the two
positive $\psi _{+}$ peaks in Figure \ref{fig-Mechanism}b with $\lambda =2.5$
and $g=1.8g_{\mathrm{s}}$. It should be noted here the adjusting of
peak positions has to pay the price to reduce the contribution of $%
E_{\Omega }$, thus requiring a large $\lambda $ and a considerable strength
of $g$ (as $g_{y}$ is proportional to $g$). Nevertheless, that is not the
only way to enhance the contribution of $E_{g_{y}}$, another way with $%
E_{\Omega }$ less affected is to introduce a node from the infinity or more exactly the
infinity sides of the secondary peaks. Such a node entering not only reverses the
sign of $\psi _{+}$ at the regime where $\partial _{x}\psi $ originally has
the the same sign as $\psi _{+}$, as in $x>4$ regime in Figure \ref%
{fig-Mechanism}b, but also makes a quick change of $\psi _{-}$ thus
increasing the amplitude of $\partial _{x}\psi $ close to the main peak of $%
\psi _{+}$ around $x=-3$. Since the main parts of the wave function and $%
E_{\Omega }$ are little affected, such a node introduction does not change
the parity, thus needing no level crossing. This is the origin of the
unconventional topological transition without gap closing. More
interestingly, this topological transition can occur both at a weak
anisotropy ($\lambda \rightarrow 1$) and in a weak coupling ($g\rightarrow 0$%
), as indicated in Figure \ref{fig-gap-P-nz}b and Figure \ref{fig-Squeezing}%
b (dashed lines).

After this unconventional topological transition, with the strengthening of the anisotropy and the coupling the node
that enters from the infinity will move closer to the peak position to form the afore-mentioned peak meeting of $\psi _{+}(x)$ and
$\partial_x\psi _{-}(x)$ as in Figure \ref{fig-Mechanism}b. The further transitions will keep this optimized near-peak node configuration
and add new nodes around the origin $x=0$, as c) to d) in Figure \ref{fig-Wigner}, rather than from the infinity. Creation of a new node around the origin will braid $\psi _{+}(x)$ and $\psi _{-}(x)$ as from d,h) to c,g) in Figure \ref{fig-NoNode-theorem}, thus accompanied with parity reversal. States with opposite parities are different quantum states and the parity reversal means level crossing of the lowest levels, thus we have gap closing for further transitions which are then conventional TPTs.

\subsection{Mechanism for the AS/PS transitions}

Besides the adjustion of the peak positions and the introduction of the node
from the infinity, there is a third way to enhance the contribution of $%
E_{g_{y}}$ which lies in reversion of the squeezing types. As
afore-mentioned for Figure \ref{fig-Mechanism}a a weak anisotropy has an
amplitude squeezing when the $\Omega $ term is more important, however when
the $g_{y}$ term comes to play a\ more dominant role at larger $\left\vert
\lambda \right\vert $, a phase squeezing will increase the strength of $%
\partial _{x}\psi _{-}$ thus leads to the AS/PS transition as described in
Figure \ref{fig-Squeezing}a,b. However such a AS/PS transition does not
occur without the first onset of the unconventional TPT. Indeed, although
the phase squeezing can amplify the amplitude of $\partial _{x}\psi _{-},$
the roughly antisymmetric profile of $\partial _{x}\psi _{-}$ around the $%
\psi _{-}$ peak position is however canceling itself if the peak of $\psi
_{+}$ is located at the same position of $\psi _{-}$ in the absence of the
node, as one can see in Figure \ref{fig-Mechanism}a around the right dashed
line at $x\approx 2.3$. So the enhancement of $\partial _{x}\psi _{-}$ by
phase squeezing does not come to effect without a node. This situation is
changed by the appearance of the node as in Figure \ref{fig-Mechanism}b
where the node around $x=3.8$ deforms the symmetric profile of $\psi _{+}$
(blue solid line) which breaks the afore-mentioned cancelation effect of the
antisymmetric $\partial _{x}\psi _{-}$. Of course the AS/PS transition does
not occur immediately after the onset of the unconventional TPT, since the
node needs to come from the infinity close enough to the peak to overcome
the amplitude squeezing caused by the $\Omega $ term. Thus, the AS/PS
transition is lagging behind the unconventional TPT. In this sense, the
AS/PS transition can be regarded as the hysteresis transition of the
unconventional TPT.

The AS/PS transition in the small-$g$ regime mentioned for Figure \ref%
{fig-Squeezing}c,d is a bit different. Reversely, rather than starting with an amplitude-squeezing state, we first show a
phase-squeezing case in Figure \ref{fig-Mechanism}c with $\lambda =5.5$ and $%
g=0.1g_{\mathrm{s}}$ in the red small-$g$ region of Figure \ref%
{fig-Squeezing}d. In this case $\psi _{+}$ and $\psi _{-}$ have a
single-polaron profile and have just been separated a little bit due to
small $g$. The profile of $\partial _{x}\psi _{-}(x)$ has two peaks with the
negative one closer to the peak of $\psi _{+}$. Note with the large $\lambda
$ here it is also favorable to get more contribution from the $g_{y}$ term
which can be enhanced by driving the negative $\partial _{x}\psi $ peak
closer to the origin. With little price to change the $\Omega $ term in the
full overlap of $\psi _{+}$ and $\psi _{-}$, this can be realized by a
larger $\xi $ as we can see from the Gaussian wave packet (\ref%
{Gaussian-polaron}) that has a derivative-peak distance from its own peak
position%
\begin{equation}
d_{\varphi ^{\prime }}=\frac{1}{\sqrt{\xi }}.  \label{d-derivative}
\end{equation}%
Thus a phase-squeezing effect is seen here. It should be mentioned this case
has a node but too weak to have a considerable effect until the PS/AS
transition. After the PS/AS transition we show a case in Figure \ref{fig-Mechanism}d with
$\lambda =3.0$ and $g=0.8g_{\mathrm{s}}$ with
an amplitude squeezing in the blue region of Figure \ref{fig-Squeezing}d. In
this case the strength of $g$ is able to separate the $\psi _{+}$ and $\psi
_{-}$ but still not enough to make the $g_{y}$ contribution overwhelming
over the $\Omega $ term, consequently the negative peak of $\partial
_{x}\psi _{-}(x)$ is located between the peaks of $\psi _{+}$ and $\psi _{-}$
as marked by the dashed line in Figure \ref{fig-Mechanism}d. On the other
hand, now the node is more visible but not yet fully reached the positive
peak of $\partial _{x}\psi $. As indicated by (\ref{d-derivative}) a smaller
$\xi $ will push the negative $\partial _{x}\psi _{-}(x)$ peak farther from
the peak of $\psi _{-}(x)$ but closer to the peak of $\psi _{+}(x)$, on the
other hand it also drives the positive $\partial _{x}\psi _{-}(x)$ peak
closer to the negative part of $\psi _{+}(x)$ beyond the node. This
process enhances the contribution of the $g_{y}$ term, while it is also
favorable for the wave-packet overlap extension in the $\Omega $ term. Thus,
a PS/AS transition also occurs in the regime with large $\lambda $ and small
$g$. Here we see the node enhancement is also a key driving factor for this
PS/AS transition.

%%%%%%%%%%%%%%%%%%%%%%%%%%%%%%%%%%%%%%%%%%%%%%%%%%%%%%%%%%%%%%%%%%%%%%%%%%%%%%%%%%%%%%%%%%%%%%%%%%
\begin{figure}[t]
\centering
\includegraphics[width=1.0\columnwidth]{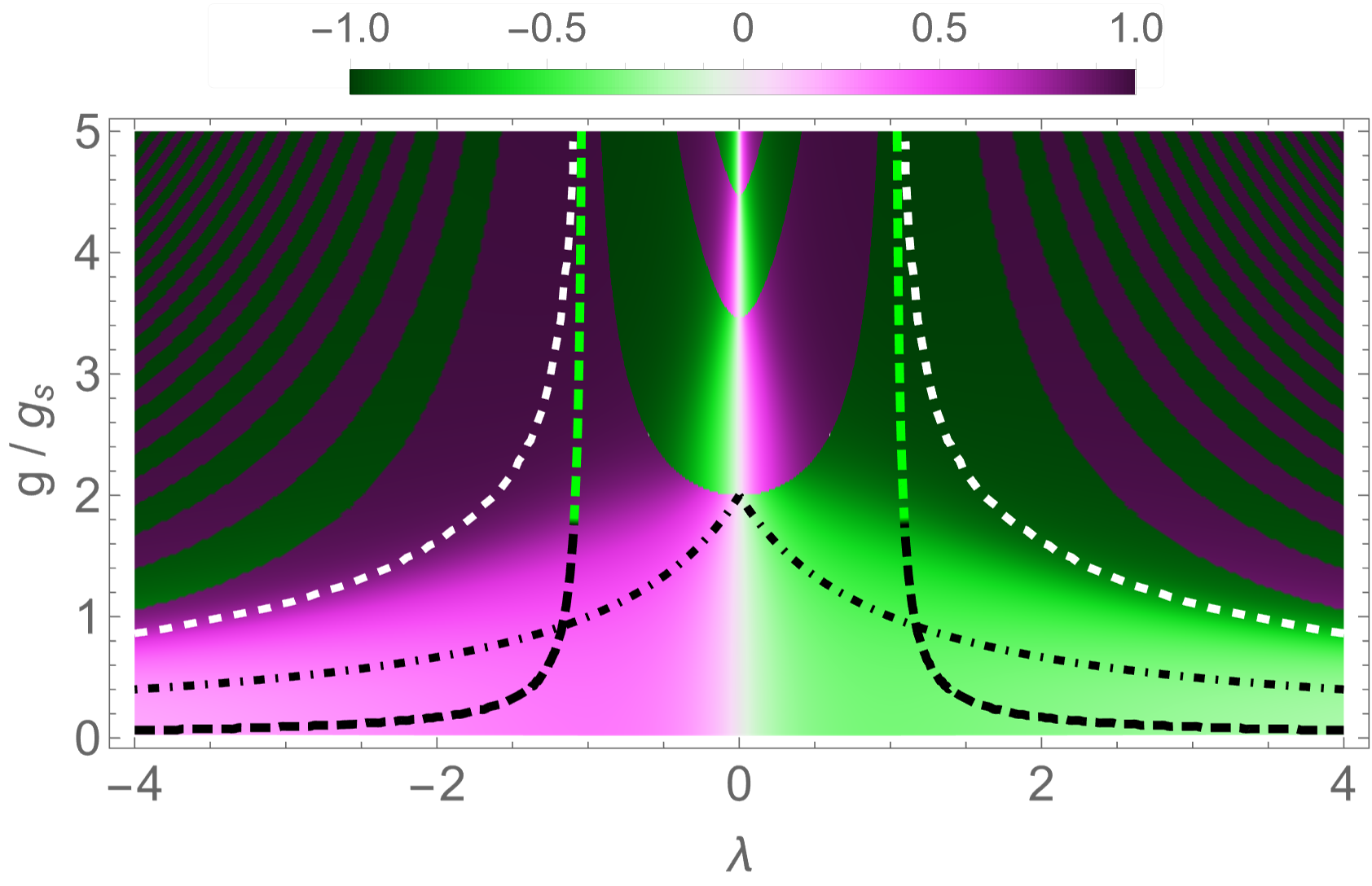}
\caption{\textit{An overview of the phase diagram: A mini-world of phase
transitions.} Density plot of $A*P$ for the GS in $\protect\lambda$%
-$g$ plane at $\protect\omega =0.5\Omega$. Here $P$ is the parity and $%
A=\langle a^{\dagger}a^{\dagger}\rangle/A_0$ scaled by $A_0=[(1+|\protect%
\lambda|)g/(2\protect\omega)]^2$. The dashed line marks the unconventional
topological transition without gap closing. The dotted line is the AS/PS
transition boundary. The dot-dashed line represents $g^{\protect\lambda}_c$
with a hidden symmetry breaking.\protect\cite{Ying-2021-AQT} The other
boundaries are locations of conventional topological transitions with gap
closing.}
\label{fig-Full-PhaseDiagram}
\end{figure}
%%%%%%%%%%%%%%%%%%%%%%%%%%%%%%%%%%%%%%%%%%%%%%%%%%%%%%%%%%%%%%%%%%%%%%%%%%%%%%%%%%%%%%%%%%%%%%%%%%

\section{Overview of Phase Diagram: a Mini-World of Phase Transitions}

\label{Sect-Overall-View}

So far we have analyzed by standing in the positive-$\lambda $ regime, the
analysis is similar for the negative-$\lambda $ regime by the mapping to
momentum space in (\ref{duality-exchange}). A panorama over full parameter
space can be obtained by the phase diagram of $\langle a^{\dagger
}a^{\dagger }\rangle =(\langle \hat{x}^{2}\rangle -\langle \hat{p}%
^{2}\rangle )/2$ multiplied by the parity $P$, as shown in \textbf{Figure %
\ref{fig-Full-PhaseDiagram}}. The parity is symmetric with respect to the
sign reversal of $\lambda $, while $\langle a^{\dagger }a^{\dagger }\rangle $
is antisymmetric as $\langle \hat{x}^{2}\rangle $ is dominant over $\langle
\hat{p}^{2}\rangle $ in the positive-$\lambda $ regime but it is reverse in
the negative-$\lambda $ regime.

The dot-dashed line is the second-order
boundary $g_{c}^{\lambda }=\frac{2}{1+\left\vert \lambda \right\vert }g_{%
\mathrm{s}}$ of the quantum phase transition in the low frequency limit,\cite%
{LiuM2017PRL} including the transition $g_{c}=g_{\mathrm{s}}$ in the QRM
with $\lambda =1$.\cite{Ashhab2013,Ying2015,Hwang2015PRL} This transition is
Landau class of phase transition with a hidden symmetry breaking despite
that the parity symmetry is preserved.\cite{Ying-2021-AQT} Meeting with this Landau class of phase transition boundary
at the hexacritical point $\left\{ \lambda ,g\right\} =\left\{ 0,2g_{\mathrm{%
s}}\right\} $ is a topological class of phase transition boundary $g_{%
\mathrm{T1}}=2g_{\mathrm{s}}/\sqrt{1-\lambda ^{2}}$ without symmetry
breaking.\cite{Ying-2021-AQT} Following the hexacritical point are a series
of quadruple points at larger $g$, formed by the series of
symmetry-protected topological transitions crossing with the
symmetry-breaking boundary at the JCM line $\lambda =0$. Bridging the QRM
and the JCM are the criticality universality in the low frequency limit\cite%
{LiuM2017PRL} and the topological universality classification at finite
frequencies reformed among the diversity and the breakdown of the
criticality universality.\cite{Ying-2021-AQT} These topological transitions
in the $\left\vert \lambda \right\vert <1$ regime are conventional type of
TPTs with gap closing, here being of first order.

The stripes of phases in the $\left\vert \lambda \right\vert >1$ regime are
also adjoined by conventional
TPTs with gap closing. The dashed lines are boundaries for the unconventional
TPT without gap closing, which are plotted in different colors in large- and
small-$g$ regimes to enhance the visibility. This unconventional TPT is of
infinite order. Lagging behind the unconventional TPT and also in the gapped
phase is the AS/PS transition with the white dashed boundaries.

At a final
glance, it seems a bit surprising that such a single-qubit system actually
opens a mini-world of phase transitions full of various ingredients: (i)
different orders of transitions, (ii) multicritical point and multiple
points, (iii) symmetry-breaking quantum phase transitions (Landau class) and
symmetry-protected topological phase transitions (topological class), (iv)
criticality universality and topological universality, (v) conventional TPTs
with gap closing and unconventional TPTs without gap closing, (v)
transitions of amplitude squeezing and phase squeezing. As a Chinese
saying goes: the sparrow may be small but it has all the vital organs.
Through this abundant mini-world of phase transitions we might gain some
deeper insights for the role of the counter-rotating term of
light-matter interactions.

\section{Conclusions and discussions}

\label{Sect-Conclusions}

We have extended the no-node theorem of the spinless-particle systems to the
fundamental QRM which involves spin coupling via light-matter interaction.
When the node number of the wave function characterizes the topological
difference of the ground state of the single-qubit system and provides a
topological classification, we have shown that the limitation of the no-node
theorem can be broken not only in the regime of coupling anisotropy between
the QRM and the JCM but also in the regime beyond the QRM. We have obtained
a full phase diagram in the interplay of the coupling and the anisotropy, which
compares the phase transitions identified by the changes of parity, the gap
closing and the jumps of the node number. Most topological transitions in
variations of the node number are accompanied with a parity reversal and a
gap closing as in the conventional TPTs, whereas we have revealed a hidden
transition in the node number without parity change or gap closing which is
an analog of the unconventional TPT without gap closing. Our mechanism
analysis shows that the conventional transitions occur with the nodes
emerging at the origin while the unconventional transition
happens with the node entering from the infinity.

We have also unveiled a transition of amplitude squeezing (AS) and phase
squeezing (PS) by tracking the frequency renormalization of the main
wave-function peak and the variance/quantum fluctuation of the
momentum. Such an AS/PS transition also occurs without gap closing but
lagging behind the unconventional TPT. The mechanism analysis indicates that
the AS/PS transition can be regarded as a hysteresis sign of the
unconventional TPT.

Both the nodal status and the squeezing effect may leave imprints in the
Wigner function. Apart from the shape deformation of the individual wave
packets, the squeezing difference of different wave packets can lead to a
local curvature of the interference fringe alignment. In particular, while
the nodeless state has always positive central interference fringes, the
node can induce a negative central interference fringe as well as an
additional negative central interference spot around zero momentum in the
phase space.

Note the anisotropic QRM is realistic and can be implemented in experimental setups of
superconducting circuits.\cite%
{Forn-Diaz2010,Pietikainen2017,Yimin2018} It may be worthwhile to stress
that the hidden unconventional TPT can be triggered in all regimes of
interaction strengths including small couplings, which would provide a great
flexibility for experimental accessibility. The gapful
situation of the unconventional TPT as well as the AS/PS transition might
also be more favorable for the condition in quantum
information processing. For an example, experimentally it is easier to cool
the system down to a gapped ground state. On the other hand, the time to
adiabatically prepare a quantum state in quantum metrology is inversely proportional to
the gap,\cite{Garbe2020} thus a transition without gap closing would have more
advantages. Since the transformed Hamiltonian (\ref{Hx}) with Rashba
spin-orbit coupling has similarity with those in cold atoms \cite{Li2012PRL}
as well as nanowires \cite{Nagasawa2013Rings,Ying2016Ellipse} and note that
nodal status could be observed in spatial density \cite{Ying2008PRL} of
Bose-Einstein condensates,\cite{Li2012PRL} we speculate our finding and
analysis might also provide some insights for the cold-atom and nanowire
systems, which we would like to address in some other works.

\section*{Acknowledgements}

This work was supported by the National Natural Science Foundation of China
(Grant No. 11974151).

%\newpage %\begin{widetext}

\appendix

\section{Proof for $E_{\Omega }^{\Phi }<E_{\Omega }^{\Psi }$ for $x_{0}=0$}

\label{Appendix-no-node}

When the node of $\Psi $ is located away from the origin, after removing the
node by the deformation to nodelss state $\Phi $ the energy reduction in the
$\Omega $ term of the QRM is obviously finite as we have demonstrated in
Figure \ref{fig-NoNode-theorem}c-f in the main text. However, when the node
is right at the origin, i.e. $x_{0}=0$ as in Figure \ref{fig-NoNode-theorem}%
g,h, the energy reduction is more delicate and needs some more rigorous
proof as presented here. The tunneling energy before and after the
deformation respectively are
\begin{eqnarray}
E_{\Omega }^{\Psi } &=&\Omega \int\limits_{-\infty }^{\infty }\Psi
_{+}(x)\Psi _{+}(-x)dx, \\
E_{\Omega }^{\Phi } &=&\Omega \int\limits_{-\infty }^{\infty }\Phi
_{+}(x)\Phi _{+}(-x)dx.
\end{eqnarray}%
Since the node is at the origin, we have $\Psi _{\pm }(x)\approx \mp kx$
around the node and $\Phi _{\pm }(x)=\pm Nk\epsilon $ for $\left\vert
x\right\vert \leqslant \epsilon $ while $\Phi _{\pm }(x)=\pm N\left\vert
\Psi _{\pm }(x)\right\vert $ for $\left\vert x\right\vert >\epsilon $, with
the renormalization factor $N=1/\sqrt{1+4k^{2}\epsilon ^{2}/3}$. Therefore,
\begin{eqnarray}
E_{\Omega }^{\Phi } &=&\Omega \left( \int\limits_{-\infty }^{\infty
}-\int\limits_{-\epsilon }^{\epsilon }\right) N^{2}\Psi _{+}(x)\Psi
_{+}(-x)dx-\Omega \int\limits_{-\epsilon }^{\epsilon }N^{2}\left( k\epsilon
\right) ^{2}dx  \nonumber \\
&=&N^{2}E_{\Omega }^{\Psi }-N^{2}\Omega \frac{4}{3}k^{2}\epsilon ^{3}
\end{eqnarray}%
so that the difference is in an order of $\epsilon ^{3}$:
\begin{eqnarray}
&&E_{\Omega }^{\Phi }-E_{\Omega }^{\Psi }  \nonumber \\
&=&\left( N^{2}-1\right) E_{\Omega }^{\Psi }-N^{2}\Omega \frac{4}{3}%
k^{2}\epsilon ^{3} \\
&=&\Omega \frac{4}{3}k^{2}\epsilon ^{3}\left\vert \int\limits_{-\infty
}^{\infty }\Psi _{+}(x)\Psi _{+}(-x)dx\right\vert -\Omega \frac{4}{3}%
k^{2}\epsilon ^{3}+O(\epsilon ^{6})  \nonumber \\
&=&-\Omega \frac{4}{3}k^{2}\epsilon ^{3}\left( 1-\left\vert
\int\limits_{-\infty }^{\infty }\Psi _{+}(x)\Psi _{+}(-x)dx\right\vert
\right) +O(\epsilon ^{6})  \nonumber \\
&=&-\Omega \frac{4}{3}k^{2}\epsilon ^{3}\Delta _{\rho }+O(\epsilon ^{6})
\end{eqnarray}%
where $\Delta _{\rho }$ is positive and finite%
\begin{eqnarray}
\Delta _{\rho } &=&\left\vert \sum_{n=0}^{\infty }\left\vert
C_{n}\right\vert ^{2}\right\vert -\left\vert \sum_{n=0}^{\infty
}(-1)^{n}\left\vert C_{n}\right\vert ^{2}\right\vert  \nonumber \\
&=&\left\vert \sum_{k=0}^{\infty }\left( \left\vert C_{2k}\right\vert
^{2}+\left\vert C_{2k+1}\right\vert ^{2}\right) \right\vert -\left\vert
\sum_{k=0}^{\infty }\left( \left\vert C_{2k}\right\vert ^{2}-\left\vert
C_{2k+1}\right\vert ^{2}\right) \right\vert  \nonumber \\
&>&0.
\end{eqnarray}%
Note here we have applied the expansion $\Psi _{+}(x)=\sum_{n=0}^{\infty
}C_{n}\phi _{n}(x)$ on the basis of quantum harmonic oscillator $\phi
_{n}(x) $ which gives\cite{Ying2020-nonlinear-bias}
\begin{equation}
\Psi _{+}(-x)=\sum_{n=0}^{\infty }C_{n}(-1)^{n}\phi
_{n}(x).
\end{equation}%
Despite the small order, the
deformation in removing the node always contributes an energy reduction in
the $\Omega $ term.

\section{Polaron Parameters in Figure \protect\ref{fig-Wigner}}

\label{Appendix-polaron-parameters}

The polaron parameters in Figure \ref{fig-Wigner} are:
a) \{$x_{\alpha }$, $x_{\beta }$, $w_{\beta }/w_{\alpha }$, $\xi _{\alpha }$, $\xi _{\beta }$\} $
= $ \{$-2.45$, $1.94$, $0.094$, $0.97$, $0.80$\},
b) \{$x_{\alpha }$, $x_{\beta }$, $w_{\beta }/w_{\alpha }$, $\xi _{\alpha }$, $\xi _{\beta }$\} $
= $ \{$-2.16 $, $2.47$, $-0.098$, $0.98$, $1.21$\},
c) \{$x_{\alpha }$, $x_{\beta }$, $x_{\gamma }$, $w_{\beta }/w_{\alpha }$, $w_{\gamma }/w_{\alpha
}$, $\xi _{\alpha }$, $\xi _{\beta }$, $\xi _{\gamma }$\} $
=$ \{$-3.11$, $2.2$, $3.75$, $0.105$, $-0.021$, $1.07$, $1.25$, $1.3$\},
d) \{$x_{\alpha }$, $x_{\beta } $, $x_{\gamma }$, $w_{\beta }/w_{\alpha }$, $w_{\gamma
}/w_{\alpha }$, $\xi _{\alpha }$, $\xi _{\beta }$, $\xi _{\gamma }$\} $
=$ \{$-2.91$, $2.0$, $3.5$, $-0.155$, $0.045$, $1.2$, $1.15$, $1.3$\}.

%%%%%%%%%%%%%%%%%%%%%%%%%%%%%%%%%%%%%%%%%%%%%%%%%%%%%%%%%%%%%
%\end{widetext}

\end{document}